# Frequency-Based Alignment of EEG and Audio Signals Using Contrastive Learning and SincNet for Auditory Attention Detection

Liao Yuan, Yuhong Zhang, Qiushi Han, Yuhang Yang, Weiwei Ding,
Yuzhe Gu, Hengxin Yang, and Liya Huang*

*Abstract*—Humans exhibit a remarkable ability to focus auditory attention in complex acoustic environments, such as cocktail parties. Auditory attention detection (AAD) aims to identify the attended speaker by analyzing brain signals, such as electroencephalography (EEG) data. Existing AAD algorithms often leverage deep learning's powerful nonlinear modeling capabilities, few consider the neural mechanisms underlying auditory processing in the brain. In this paper, we propose SincAlignNet, a novel network based on an improved SincNet and contrastive learning, designed to align audio and EEG features for auditory attention detection. The SincNet component simulates the brain's processing of audio during auditory attention, while contrastive learning guides the model to learn the relationship between EEG signals and attended speech. During inference, we calculate the cosine similarity between EEG and audio features and also explore direct inference of the attended speaker using EEG data. Cross-trial evaluations results demonstrate that SincAlignNet outperforms state-of-the-art AAD methods on two publicly available datasets, KUL and DTU, achieving average accuracies of 78.3% and 92.2%, respectively, with a 1-second decision window. The model exhibits strong interpretability, revealing that the left and right temporal lobes are more active during both male and female speaker scenarios. Furthermore, we found that using data from only six electrodes near the temporal lobes maintains similar or even better performance compared to using 64 electrodes. These findings indicate that efficient low-density EEG online decoding is achievable, marking an important step toward the practical implementation of neuro-guided hearing aids in real-world applications. Code is available at: https://github.com/LiaoEuan/SincAlignNet.

## I. Introduction

IN complex auditory environments, such as a "cocktail party", multiple speakers and background noises coexist, individuals with normal hearing can easily focus on and selectively attend to a target speaker. This ability is known as selective auditory attention. However, for individuals with hearing impairments, achieving the same level of focus is challenging. Current hearing aids typically assist by reducing background noise and enhancing speech clarity[1]. Yet, these devices lack the capability to determine which sound source the user wishes to focus on. Specifically, they cannot identify which speech signals (i.e., the attended speaker) to enhance and which signals to suppress as noise (i.e., the unattended speakers). We refer to this issue as the auditory attention decoding (AAD) problem.

Recent studies have demonstrated that auditory attention can be decoded from brain activity recordings, such as electrocorticography (ECoG) [2], magnetoencephalography (MEG) [3], [4], and electroencephalography (EEG) [5], [6], [7], in multi-speaker environments. Decoding auditory attention opens up numerous possibilities for human-machine interaction systems, such as cognitive control in hearing aids [8] (i.e., neuro-guided hearing aids) and rehabilitation through neurofeedback. Since EEG provides a non-invasive method for investigating cortical activity with high temporal resolution [9], it has become particularly valuable for decoding auditory attention in brain-computer interface applications. This paper focuses specifically on decoding auditory attention from EEG signals.

Early EEG-based AAD methods primarily relied on linear models, which can be classified into backward and forward models. The backward model involves reconstructing the auditory stimulus from EEG signals and then computing the correlation between the reconstructed stimulus and the speech envelope of the attended speaker [10]. A prominent example of this approach is the canonical correlation analysis (CCA) model [11], [12]. The CCA-based model distinguishes the attended speaker by comparing the Pearson correlation coefficient between the reconstructed stimulus and the speech envelope of two competing speakers. The forward model, also known as the temporal response function (TRF) approach [13], uses linear system-response models to study continuous speech stimuli

This work was supported by the National Natural Science Foundation of China (Grant No. 61977039). (Yuan Liao and Yuhong Zhang contributed equally to this work.) (Corresponding author: Liya Huang).
Yuan Liao, Qiushi Han, Yuhang Yang, Yuzhe Gu, Hengxi Yang, and Liya Huang are with college of electronic and optical engineering & college of flexible electronics, Nanjing University of Posts and Telecommunications, Jiangsu, 210023, China, E-mail: {1022020619; huangly}@njupt.edu.cn.

Yuhong Zhang is with the Department of Bioengineering, University of California, San Diego, La Jolla, 92093, USA, E-mail: yuz291@ucsd.edu. Weiwei Ding is with the Department of Biomedical Engineering, BeiHang University, BeiJing, China, E-mail: weiweiding@buaa.edu.cn.

through estimated TRFs. These methods have proven effective for examining how cortical processing of speech features is modulated by selective auditory attention. Specifically, TRFs corresponding to the attended speaker exhibit pronounced peaks around 100 and 200 ms, which are less prominent in TRFs for the ignored speaker [14], [15]. However, linear models typically require a longer decision window to extract brain activity features, often around 10 seconds, and their performance quickly deteriorates as the decision window narrows [16], [17].

It is well established that the human auditory system is inherently nonlinear [18]. Since neural networks are capable of nonlinear modeling, several studies have proposed models based on convolutional neural networks (CNNs). These models use CNN-based stimulus reconstruction algorithms to infer attention [19], [20]. Other approaches [21], [22] have directly classified attention by predicting which speaker (Speaker 1 or Speaker 2) the listener is focusing on from EEG signals. Building on this foundation, Paper [23] employed Long Short-Term Memory (LSTM) networks to capture dependencies within EEG signals, achieving commendable decoding performance. Recognizing the importance of spatial distribution in EEG data, the study [24] introduced DARNet, a dual-attention refinement network that integrates temporal and spatial constructs. DARNet extracts temporal patterns at various levels within EEG signals, enhancing the model's ability to capture long-term dependencies and thereby improving performance. Advancing further, more sophisticated methods such as end-to-end models have been developed to leverage EEG for speech separation, facilitating the extraction of attended audio. One notable example is NeuroHeed [25], which employs a learned mask to simulate how the brain selectively focuses on certain audio while filtering out the rest. While these models benefit from deep learning's robust fitting capabilities, they often involve highly complex processes, making the interpretation of underlying parameters challenging. As a result, only a few models provide significant insights into the neural mechanisms or the neuroscience underpinning auditory attention.

As we know, humans excel at flexible and effective auditory attention selection—an ability that current intelligent machines cannot replicate. This motivates us to transfer human auditory attention capabilities to models by aligning them with brain representations, which could help bridge the gap in auditory perception and improve AAD performance. In the human auditory pathway, speech is broken down into a series of band-pass filtered signals, and the rich representation of speech is processed as it reaches the auditory cortex [26], [27]. Inspired by these findings, we aim to explore a model that dynamically aligns EEG and speech signals across different frequency bands based on their interactions, in order to simulate the top-down and bottom-up regulation of auditory attention [28], thereby enhancing AAD performance.

To achieve this, we propose a contrastive learning approach called SincAlignNet, which directly learns the correspondence between EEG and audio by maximizing the mutual information between the encoding of each EEG segment and the encoding of the attended audio. We utilize and improve upon SincNet as the base encoder for both EEG and audio. SincNet [29] is a specialized one-dimensional CNN with band-pass filtering capabilities, enabling it to learn features within a specific frequency range. Our approach focuses on directly learning the relationship between the brain and speech from a frequency perspective. The encoder module, guided by contrastive learning, determines how to emphasize or suppress various frequency components for both EEG and speeches. Ideally, contrastive learning will reveal a common feature space for EEG and audio, where each EEG encoding is positioned closest to the corresponding attended audio encoding, allowing for accurate inference of the attended audio corresponding to a given EEG segment.

To assess the performance of SincAlignNet, we evaluated its decoding accuracy on two datasets: KUL[5] and DTU[7]. The main contributions of this paper are summarized as follows:

1) We introduce the SincAlignNet model, which simulates human auditory attention behavior. By using contrastive learning to align EEG and audio modalities, it dynamically captures the interaction between brain responses and speech stimuli, achieving the highest accuracy compared to existing models.
2) Our model is capable of detecting auditory attention activity in the brain. Experimental results show that the alpha band in the temporal lobe is most active during auditory attention, which aligns with existing neuroscience research.
3) We demonstrate that using just six electrodes around the temporal lobe yields similar or even better performance than using 64 electrodes. Moreover, the model's complexity is relatively low, with only 0.34 million parameters, making it promising for future neuro-guided hearing aid development.

The paper is structured as follows: Section II provides a detailed description of the proposed method. Section III presents our experimental setup and results. Section IV discusses the strengths and weaknesses of our approach. Finally, Section V concludes the paper and suggests potential avenues for future research.

## II. METHODS

This section introduces the SincAlignNet framework, illustrated in Fig. 1. SincAlignNet integrates two main components—contrastive learning and inference—which are described in the following subsections.

Given N pairs of EEG and audio data as input, SincAlignNet first extracts features from both modalities using SincNet-based encoders. During the contrastive learning phase, the system projects these features into a shared latent space. Within this space, each EEG representation is drawn closer to the representation of its corresponding attended audio while being separated from the remaining EEG and audio representations. Once EEG and audio features are aligned, the framework employs two inference methods to determine which audio signal the participant is focusing on.

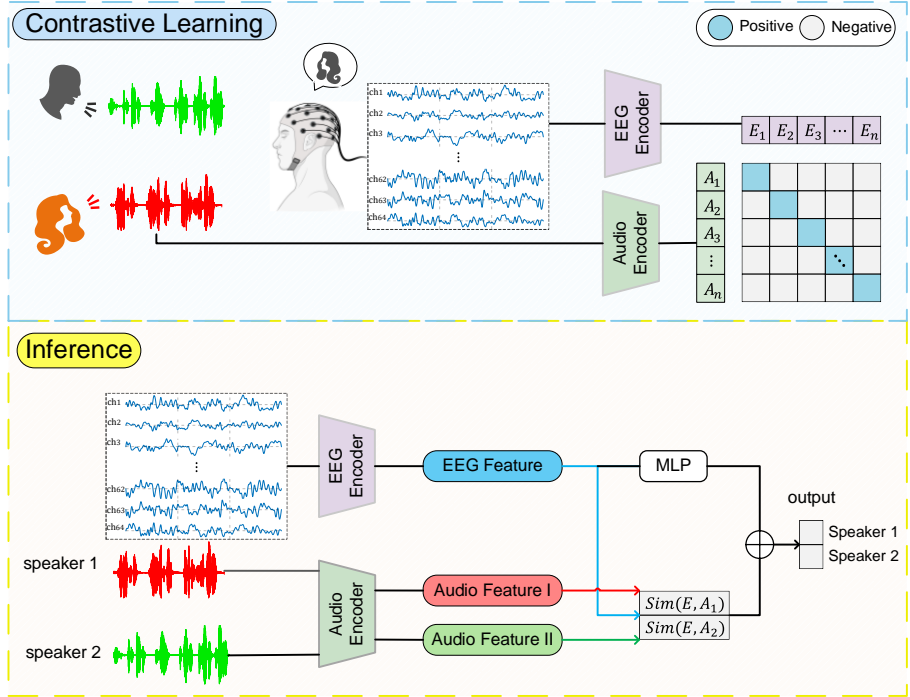

Fig. 1. The Framework of the SincAlignNet Model for AAD, which mainly consists of two parts: Contrastive Learning and Inference. Contrastive learning aligns EEG encoding with attended audio encoding by maximizing the mutual information of correct EEG-Audio pairs. Inference is used to identify the audio that the participant is attending to, by calculating the cosine similarity between EEG features and audio features, and also considering the use of EEG features for direct inference of the attended speaker.

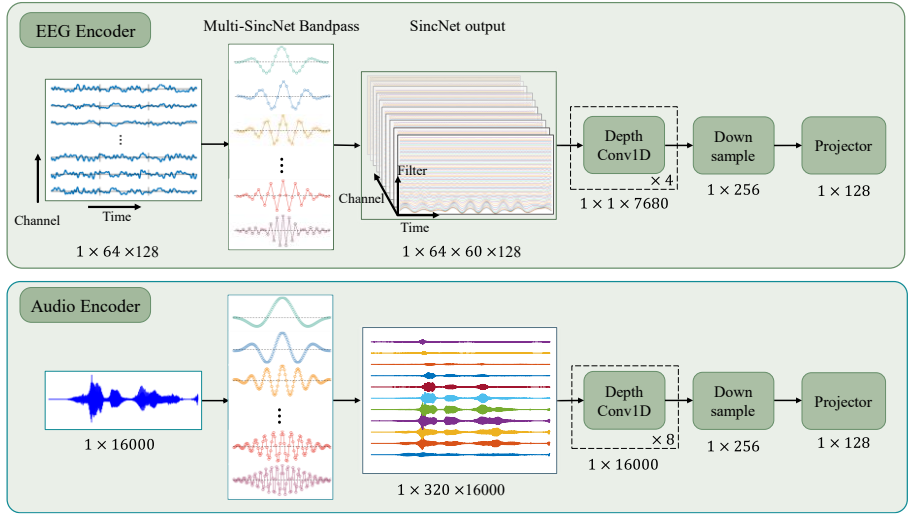

Fig. 2. Details of the EEG encoder and Audio Encoder. Both the encoders consist of four main components: Multi-SincNet Bandpass, Depth Conv1D, Down Sample, and Projector. Initially, the input signal is processed by the SincNet Bandpass filter, which applies 60 filters for the EEG encoder and 320 filters for the audio encoder. Next, Depth Conv1D combines the outputs from these filters to extract deeper features. After that, the signal is compressed using a Down Sample module to reduce the data dimension while preserving key information. Finally, the Projector maps the data into a 128-dimensional feature space.

### A. Problem Definition

We now briefly introduce the experimental process of auditory attention detection and speech selection based on EEG data. Let the paired dataset $\Omega = \{(A_1^+, A_2^-, E)\}$, where $A_1^+$ and $A_2^-$ represent the attended and ignored audio signals respectively, and $E$ represents the corresponding EEG signals recorded in synchronization with the audio stimuli. Specifically, $A_1^+, A_2^- \in R^{B \times L}$, with $B$ being the batch size and $L$ is the temporal length of the Audio. Similarly, $E \in R^{B \times C \times T}$, where $C$ is the number of EEG channels capturing the participant's brain activity and $T$ is the temporal length of the EEG recording.

During each experiment, participants are instructed to focus on one of the audio signals while ignoring the other. The objective is to develop a model that, given the audio signals $A_1$ and $A_2$ and the corresponding EEG data $E$, can accurately identify which audio signal the participant is focusing on.

## B. Model Assumptions

To emulate the human brain's flexible auditory attention selection capabilities, we propose a frequency-aligned contrastive learning paradigm based on the following assumptions:

1) As illustrated in Fig. 3 (a), the brain receives mixed audio signals and is capable of performing noise reduction to ultimately obtain relatively clear audio of the attended speaker. We hypothesize that this noise reduction process can be modeled using the SincNet architecture to simulate the filtering process.

2) In noisy environments, when humans focus their attention on a single speaker, we assume that, from an information entropy perspective, the brain's processing of the heard audio minimizes mutual information entropy as much as possible. We simulate this process using the contrastive learning approach depicted in Fig. 3 (b).

Building on the two assumptions, we will further elaborate on the proposed SincAlignNet model in this paper.

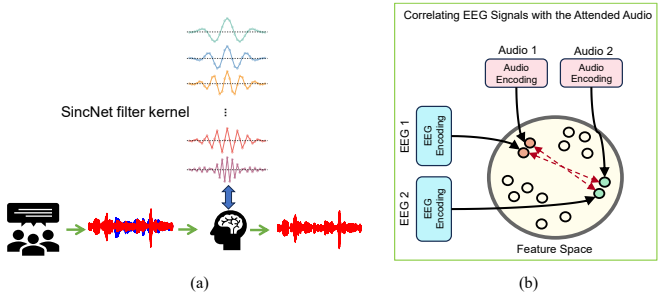

Fig. 3. Illustration of Model Assumptions. (a) Noise Reduction Process using SincNet; (b) Contrastive Learning Approach for Minimizing Mutual Information Entropy.

## C. SincNet Based Encoder

SincNet is a specialized one-dimensional CNN with bandpass filtering capabilities, enabling feature learning within a specific frequency range. In this study, it serves as the primary encoder for both EEG and audio signals, positioned closest to the input data.

In the frequency domain, the magnitude of a bandpass filter can be expressed as the difference between two low-pass filters:

$$G[f, f_1, f_2] = rect\left(\frac{f}{2f_2}\right) - rect\left(\frac{f}{2f_1}\right) \quad (1)$$

where $f_1$ and $f_2$ are the learned low and high cutoff frequencies, and rect(.) is the rectangular function in the magnitude frequency domain. Transforming this into the time domain, the corresponding function $g$ is given by:

$$g[n, f_1, f_2] = 2f_2 sinc(2\pi f_2 n) - 2f_1 sinc(2\pi f_1 n) \quad (2)$$

where the sinc function is defined as $sinc(x) = \frac{sin(x)}{x}$.

By using the Hamming window function to improve the edge processing of the signal, the signal becomes smoother during frequency domain analysis. The optimized $g_w$ is:

$$g_w[n, f_1, f_2] = g[n, f_1, f_2] \cdot w[n] \quad (3)$$

where the Hamming window $w[n]$ is defined as:

$$w[n] = 0.54 - 0.46 \cdot cos\left(\frac{2\pi n}{L}\right) \quad (4)$$

This enhancement reduces spectral leakage and ensures smoother transitions during frequency domain analysis.

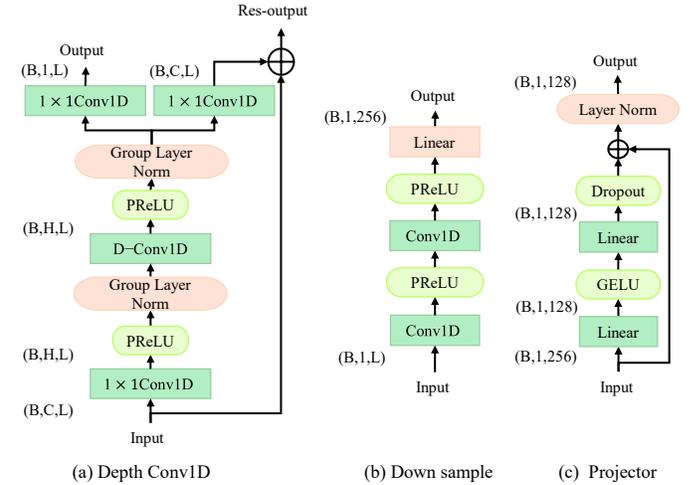

Fig. 4. Details of each module. (a) Depth-wise 1D convolution block. (b) Down sample module. (c) Projector.

## D. EEG Embedding and Audio Embedding

To ensure alignment between the EEG and audio modalities, we designed their encoder architectures to be highly similar. Both the EEG and Audio Encoders consist of four key components: SincNet Bandpass, Depth Conv1D, Down Sampling, and Projector.

As illustrated in Fig.2, The input signal first passes through the SincNet Bandpass filter to extract features from specific frequency ranges. For the EEG Encoder, 60 filters are employed to capture the unique characteristics of EEG data, while the Audio Encoder uses 320 filters to account for the greater complexity of audio signals. Following this, depth-wise 1D convolutional layers process the outputs from the filters, combining sub-band features to extract deeper patterns and enhance representation quality. Next, a down sampling module is applied to compress the feature dimensions, retaining key information while improving computational efficiency. Finally, the Projector module maps the features into a shared 128-dimensional space, which is essential for enabling contrastive learning and ensuring effective comparisons between the EEG and audio modalities. This consistent architecture ensures compatibility and facilitates robust feature extraction and alignment across both signal types.

1) Depth-wise convolution blocks: As illustrated in Fig.4 (a), Depth-wise convolution block [30] decouples the standard convolution operation into two consecutive operations: a Depth-wise convolution followed by a pointwise convolution. This approach significantly reduces the model size. Each DepthConv1D module comprises multiple layers, with residual connections in every layer to ensure the generation of multi-level feature outputs.

Within N layers Depth-wise convolution stack, the layers employ depth-wise convolutions with a dilation factor that increases exponentially, $2^d$, where $d \in \{0, \cdots, N-1\}$. The residual output from the previous DepthConv1D layer serves as 4the input for the next Depth Conv1D layer, with only the output from the final DepthConv1D layer being considered as the overall output.

2) Down sample Convolution Module: As shown in Fig.4 (b), this module compresses the data using two standard 1D convolutional layers. Both layers operate on single-channel data, with kernel sizes of 64 and 32 and strides of 6 and 4, respectively. Between these layers, a PReLU activation function introduces non-linear mapping. The module's final output passes through a linear layer that projects the features into a 256-dimensional space, efficiently reducing data dimensionality while retaining essential information.

3) Projector Module: The Projector Module maps feature into a 128-dimensional space optimized for subsequent contrastive learning tasks, as depicted in Fig.4 (c). It consists of a linear layer, a GELU activation function, a dropout layer, and a layer normalization step. This module reduces the 256-dimensional feature embedding to 128 dimensions, ensuring more discriminative feature representations. The dropout layer, active during training, minimizes the risk of overfitting and improves generalization.

Finally, the EEG Embedding $Z_{E^i}$ and Audio Embedding $Z_{A^i}$ are obtained from the EEG encoder and audio encoder, respectively, expressed as follows:

$$Z_{E^i} = f_{EEG}(E^i) \tag{5}$$
$$Z_{A^i} = f_{Audio}(A^i) \tag{6}$$

### E. EEG Guidance AAD

In this study, the model consists of two phases: constructive training phase and inference phase, aimed at detecting the auditory attention of participants based on EEG signals, as illustrated in Fig.1.

1) Constructive learning phase: Inspired by CLIP [31], a recent successful cross-modal approach, we apply cross-modal contrastive learning to align EEG and audio modalities at the instance level. Given a batch of B pairs of EEG-audio tuples $\{(A_1^+, A_2^-, E)\}_{i=1:B}$, our objective is to maximize the mutual information between correctly positive paired $\{(A_1^+, E)\}_{i=1:B}$ samples. This is achieved by optimizing a $B \times B$ similarity matrix, as illustrated in Fig. 1, which represents the cross-modal pairwise similarities in a shared representation space. We use cosine similarity to represent mutual information, as defined in (7). For this matrix, each column corresponds to an EEG clip, and each row corresponds to an audio signal. The (i,j)-th entry indicates the similarity between the i-th EEG clip and the j-th audio sample. The diagonal entries are maximized to reflect correct pairings, while off-diagonal entries are minimized to reduce incorrect pairings. We employ a constructive learning loss function to achieve the desired alignment, as shown in (8).

$$sim(a, b) = \frac{a^T b}{\|a\|\|b\|} \tag{7}$$

$$\mathcal{L}_{CLIP} = -\sum_{i=1}^{B} \log \frac{e^{(\frac{sim(Z_E^i, Z_{A^+}^i)}{\tau})}}{\sum_{j=1}^{N} e^{(\frac{sim(Z_E^i, Z_{A^+}^j)}{\tau})}} \tag{8}$$

2) Inference phase: To determine the attended audio source, we employ a hybrid approach that combines EEG Driven Inference (EDI) and EEG-Audio Correlation Inference (EACI), leveraging both EEG-based predictions and similarity-based comparisons for robust inference.

In the EACI process, we calculate the cosine similarity between the EEG embedding $Z_E$ and each of the two audio embeddings $Z_{A_1}$ and $Z_{A_2}$. This results in two similarity scores:

$$EACI = [sim(Z_E, Z_{A_1}), sim(Z_E, Z_{A_2})] \tag{9}$$

In parallel, the EDI component processes $Z_E$ through a Multi-Layer Perception (MLP). The MLP outputs a 2-dimensional vector $[p_1, p_2]$, where $p_1$ and $p_2$ represent the likelihood of attending to audio sources $A_1$ and $A_2$, respectively:

$$EDI = MLP(Z_E) = [p_1, p_2] \tag{10}$$

To combine these two inference methods, we sum the outputs from the EACI and EDI processes, producing a final decision output.

$$Output = EDI + EACI \tag{11}$$

The overall loss function for the model is a combination of the contrastive loss $\mathcal{L}_{CLIP}$ for the constructive learning phase and the cross-entropy loss for the inference phase $\mathcal{L}_{CE}$:

$$Loss = \mathcal{L}_{CLIP} + \mathcal{L}_{CE} \tag{12}$$

## III. EXPERIMENT AND RESULTS

### A. AAD Dataset

In this study, we conducted experiments on two publicly available Auditory Attention Detection (AAD) datasets: the KUL dataset [5] and the DTU dataset [7].

KUL Dataset: EEG data were collected from 16 participants with normal hearing. The stimulus materials consisted of four Dutch short stories narrated by different male speakers. In each trial, participants listened to two competing audio streams simultaneously in a soundproof and electromagnetically shielded room. The stimuli were delivered using Head-Related Transfer Function (HRTF) filtering or directly binaurally, with one speaker positioned at a 90-degree angle to the left and right of the subject. Participants were instructed to selectively attend to one of the audio streams. Each participant completed 8 trials, with each trial lasting 6 minutes. Overall, each participant contributed approximately 48 minutes of EEG data, resulting in a dataset comprising 12.8 hours of recorded data.

DTU Dataset: EEG data were collected from 18 participants with normal hearing, with speech stimuli presented by male and female native speakers in either a soundproof or reverberation chamber. Participants were presented with a mixture of speech, with two speech streams spatially oriented at -60° and +60° azimuth angles. The position and gender of the target speaker were randomized within the trials, ensuring an equal number of left-ear and right-ear trials. EEG data were recorded using a BioSemi Active system with 64 channels at a sampling rate of 512 Hz, following the international 10/20 electrode placement system. Each participant completed a total of 60 trials, with each trial consisting of auditory stimuli lasting 50 seconds. In total, each participant contributed 50 minutes of EEG data, leading to a cumulative total of 15 hours of EEG data from all 18 participants.

## B. Test and Validation

Based on the recommendations from [32], we adopted a cross-trial validation method to ensure that the stories used during training do not appear in the test set. For the KUL dataset, the folds are organized as (trial1, trial2), (trial3, trial4), (trial5, trial6), and (trial7, trial8). For the DTU dataset, we use the first 48 trials for training and the last 12 for testing. Hence, each trial only appears in either training or test set. The average of these results was taken as the experimental outcome for the current participant.

## C. Datasets Preprocessing

The EEG data were first preprocessed to remove 50 Hz power-line noise using a notch filter. A fourth-order Butterworth high-pass filter with a cutoff frequency of 0.1 Hz was then applied to eliminate low-frequency drifts. This filter was designed to ensure that the group delay at frequencies above 1 Hz was less than two samples, thereby preserving temporal accuracy. Subsequently, the EEG signals were re-referenced to the average response across all channels and filtered within a frequency range of 1–60 Hz using a band-pass filter. The filtered signals were then down sampled to 128 Hz to reduce data dimensionality. Independent Component Analysis (ICA) was applied to remove artifacts such as eye movement and muscle activity. Finally, Z-score normalization was performed on each trial to standardize the data and minimize individual variability.

The audio signals were down sampled to 16 kHz to reduce data volume while retaining critical audio features. Gammatone filter banks were then applied to the audio data to simulate human auditory perception of different frequency ranges. The filter bank covered a frequency range of 80 Hz to 8000 Hz, with center frequencies distributed evenly based on the Equivalent Rectangular Bandwidth (ERB) scale. This ensured effective capture of key frequency components in the audio signals. Afterward, the EEG and audio signals were temporally aligned. Finally, Z-score normalization was applied to all audio data to standardize the features and remove global biases.

## D. Network Configuration

To illustrate the network configuration in detail, we use a 1-second decision window as an example. The model input consists of 1 second of EEG signals ($E \in R^{64 \times 128}$) and two audio signals, $A_1^+$ and $A_2^-$ ($A \in R^{1 \times 16000}$). The EEG data include 64 channels and 128 sampling points, while the audio data are single-channel with 16,000 sampling points.

The cutoff frequencies of the SincNet filter bank are initialized differently for the EEG Encoder and the Audio Encoder. For the EEG Encoder, we use 60 filters with cutoff frequencies uniformly distributed on the Hz scale. For the Audio Encoder, 320 filters are employed, with cutoff frequencies uniformly distributed on the Mel scale to mimic human auditory perception. The human ear is more sensitive to low-frequency sounds, but this sensitivity diminishes gradually as frequency increases. As a result, the initialization of high and low cutoff frequencies follows a curve-like distribution.

During training, we set the number of epochs to 150 and minimized the loss function using the Adam optimizer with a learning rate of 0.0001. A dropout rate of 0.1 was applied, and the batch size was configured to 32. All models were implemented using the PyTorch framework and trained on an NVIDIA A100. Detailed model parameter initialization is provided in Table I.

Table I. Model Parameter Configuration.

| | EEG Encoder | Audio Encoder |
| --- | --- | --- |
| **Input** | $E \in R^{64 \times 128}$ | $A_1^+, A_2^- \in R^{1 \times 16000}$ |
| **SincNet** | | |
| hidden channel | 60 | 360 |
| kernel size | 31 | 101 |
| min low hz | 1 | 50 |
| min band hz | 4 | 50 |
| **Depth Conv1D** | | |
| hidden channel | 32 | 32 |
| kernel size | 3 | 3 |
| strides | [6,4] | [6,4] |
| linear | [310,256] | [491,256] |
| **Projector Module** | | |
| linear1 | [256,128] | [256,128] |
| linear2 | [128,128] | [128,128] |
| **Output** | $Z_E \in R^{1 \times 128}$ | $Z_{A_1}, Z_{A_2} \in R^{1 \times 128}$ |

## E. Decoding Performance

To test the generalization of the proposed model, we compared the performance of SincAlignNet with state-of-the-art models, including a total of four baselines. Including CNN [22], BIAnet [33], Graph-EEG Net [34], DARNet [24]. A decision window length of 1 second was selected, as it is the common decision window length used by all included baselines. Additionally, a 1-second decision window approximates the time delay required for humans to shift attention. The experimental results are shown in Table II. The results are replicated from the corresponding papers. We conducted paired t-tests on the decoding accuracy across subjects for both the KUL and DTU datasets to evaluate the performance of SincAlignNet compared to baseline models.

Table II: AAD accuracy (%) comparison on DTU and KLU dataset with 1s Decision window.

| Model | Year | Ues Auditory | KUL | DTU |
| --- | --- | --- | --- | --- |
| CNN [22] | 2021 | yes | 63.5 | 56.6 |
| BIAnet [33] | 2022 | yes | 60.9 | 59.1 |
| Graph-EEG Net[34] | 2023 | no | 56.6 | 64.3 |
| DARNet [24] | 2024 | no | 76.8 | 69.1 |
| **SincAlignNet (ours)** | 2024 | yes | 78.3 | 92.2 |

For the KUL dataset, SincAlignNet achieved an accuracy of 78.3%, slightly outperforming the advanced DARNet model with a 1.5% relative improvement. Compared to the CNN baseline, SincAlignNet demonstrated a 14.5% increase in accuracy. Paired t-tests validated these improvements as statistically significant, with a p-value < 0.001.

On the DTU dataset, SincAlignNet delivered significant improvements in decoding accuracy, achieving an accuracy of 92.2% with a 1-second decision window. Relative to the baseline models, SincAlignNet showed improvements of 33.6%, 31.1%, 25.9%, and 21.1% over CNN, BIAnet, Graph-EEG Net, and DARNet, respectively. Statistical analysis through paired t-tests confirms the significance of these improvements, with all comparisons yielding $p < 0.0001$, indicating that SincAlignNet significantly outperforms each baseline models.

*F. Effect of the Decision Window Length*

In addition, we evaluated the accuracy of SincAlignNet at different decision window lengths (0.5, 1, 1.5, 2, and 2.5 seconds) on both the KUL and DTU datasets, as shown in Fig.5. The results demonstrate that SincAlignNet achieves robust performance on the DTU dataset, with average accuracy consistently between 86% and 94%, regardless of the decision window length. The standard deviation remains relatively low, indicating strong stability across subjects. On the KUL dataset, however, the model's average accuracy ranges from 68% to 82%, with larger variability compared to the DTU dataset.

We attribute the stability observed in the DTU dataset to the fact that it includes both male and female speakers, which makes distinguishing between speakers easier due to the diversity of voice characteristics. In contrast, the KUL dataset contains only male speakers with similar voices, which likely increases the difficulty for the model in distinguishing between speakers, leading to more pronounced fluctuations in performance.

Moreover, the effect of decision window length appears to differ between the two datasets. While extending the decision window generally helps maintain high accuracy on DTU, the performance on KUL fluctuates more notably, indicating that longer time windows do not always enhance model performance in scenarios with high similarity among speakers. This suggests that future work could explore incorporating additional features or using more sophisticated attention mechanisms to better handle similar speakers.

Overall, these findings highlight SincAlignNet's robustness in diverse conditions, while also pointing to specific challenges in handling speaker homogeneity, especially under varying decision window lengths.

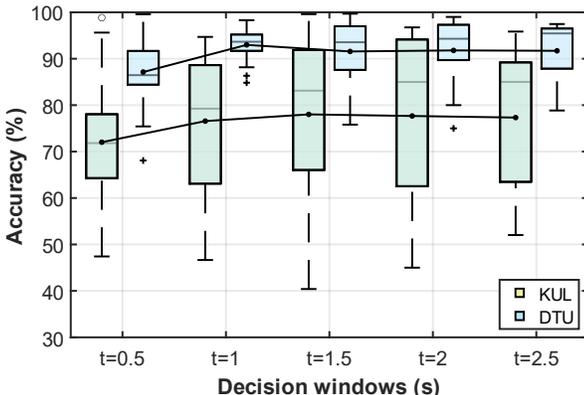

Fig. 5. AAD accuracy (%) of SincAlignNet across different decision windows (0.5, 1, 1.5, 2, 2.5 seconds) on KUL and DTU datasets.

## IV. DISCUSSION

This section first evaluates the functionality of key components in the SincAlignNet model through ablation experiments. It then analyzes the frequency features learned by the SincNet in the EEG encoder, with additional visualizations for further analysis. Finally, the model's parameter size is compared and evaluated.

*A. Ablation Experiments*

We conducted comprehensive ablation experiments to evaluate the contributions of different components in our model. Specifically, we individually removed the EDI and EACI branches while keeping all other experimental conditions consistent with previous settings. To assess the role of contrastive learning, we also used only the Inference module directly for classification without contrastive learning. In all scenarios, we ensured that the model parameters were fully optimized to guarantee peak performance, regardless of whether modules were removed or added. The results of these experiments are summarized in Table III.

For the KUL dataset, the dual inference approach incorporating both EDI and EACI yielded the best results. Removing the EDI branch had the most significant impact, leading to a performance drop of 25.6%, while the removal of the EACI branch resulted in a smaller performance loss of 7.58%. This suggests that the EDI component is crucial when speaker voices are highly similar, as it significantly contributes to distinguishing between them.

In contrast, for the DTU dataset, removing the EACI branch had the most substantial effect, causing a performance loss of 28.04%. Interestingly, removing the EDI branch in this case led to a performance improvement of 4.3%. This indicates that the EACI component is particularly useful in scenarios involving speakers of different genders, as it effectively captures the distinct features between male and female voices.

Regarding contrastive learning, we observed that excluding it led to performance declines of 12.01% on the KUL dataset and 6.65% on the DTU dataset. These findings demonstrate that contrastive learning enhances the model's ability to align EEG and audio features, thereby improving overall performance.

Table III: Ablation Study on KUL and DTU dataset with 1s Decision window.

| Dataset | Experimental Conditions | Accuracy (%) | Changes (%) |
|---|---|---|---|
| KUL | w/o EDI | 52.78 | -25.78 |
| | w/o EACI | 70.72 | -7.64 |
| | w/o contrastive learning | 66.35 | -12.01 |
| | **SincAlignNet (ours)** | **78.36** | - |
| DTU | w/o EDI | 94.50 | +2.29 |
| | w/o EACI | 62.16 | -30.05 |
| | w/o contrastive learning | 85.56 | -6.65 |
| | **SincAlignNet (ours)** | **92.21** | - |

Changes: Compared with the original SincAlignNet

*B. What did SincNet Learn?*

To elucidate the specific frequency components modulated by the SincNet EEG encoder, we conducted a comparative analysis of the magnitude frequency responses before and after training. Specifically, for models trained on the KUL and DTU datasets with a 1-second decision window, we computed the magnitude frequency response of each SincNet filter. Subsequently, we subtracted the initialized magnitude frequency response from the trained magnitude frequency response, averaged these differences across all filters, and visualized the results in the bar charts presented in Fig. 6. In these figures, positive values indicate frequency components that were enhanced post-training, while negative values signify frequencies that were suppressed. It is noteworthy that the horizontal axis displays frequency with a resolution of 4 Hz, determined by the convolution kernel size of 31 in the EEG encoder and an EEG sampling rate of 128 Hz ($128/31 \approx 4$).

In Fig. 6(a), pertaining to the DTU dataset with both male and female speakers, we observe that frequency components within the 12 to 16 Hz range are significantly enhanced. Conversely, frequencies around 8 Hz, 20 Hz, 40 Hz, and 52 Hz exhibit notable suppression. The enhancement in the beta frequency band (12-16 Hz) aligns with existing literature [35], which indicates that higher network segregation of coupled beta-band oscillations supports faster auditory perceptual decisions across trials. Additionally, the pronounced emphasis on 12 Hz is consistent with previous studies [36], [37], demonstrating that rhythmic activity in the alpha band within the auditory association cortex plays a crucial role in controlling attention to currently relevant segments within an auditory scene.

Unlike previous studies, our analysis extends the frequency range beyond 32 Hz. We found that EEG data within the 32 to 52 Hz range are relatively suppressed, suggesting that the SincNet encoder selectively attenuates higher-frequency components that may be less relevant for auditory attention tasks.

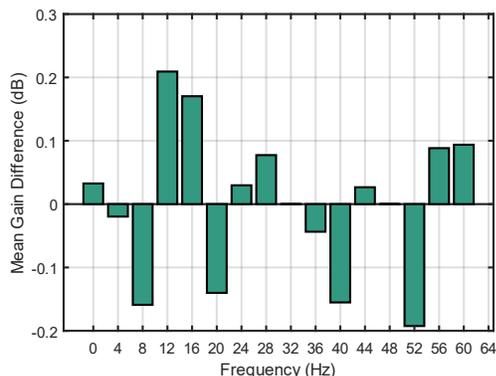

Fig. 6(a). Enhancement and Suppression of EEG Frequency Components in SincNet Filters After Training Compared to Initialization on the DTU Dataset (Male and Female Speakers). Positive values indicate enhancement, negative values indicate suppression.

In contrast, Fig. 6(b) illustrates the results for the KUL dataset, which includes two male speakers. Here, we observe a general emphasis across all EEG frequencies. Notably, similar to the DTU dataset, the 16 Hz frequency component is particularly enhanced, whereas frequencies around 40 Hz and 52 Hz are suppressed. The consistent enhancement of the 16 Hz component across both datasets reinforces the robustness of the SincNet architecture in identifying and amplifying critical frequency bands associated with auditory attention. Furthermore, the suppression of higher frequencies (40 Hz and above) in both datasets highlight the model's proficiency in isolating and prioritizing frequencies most pertinent to the cognitive processes under investigation.

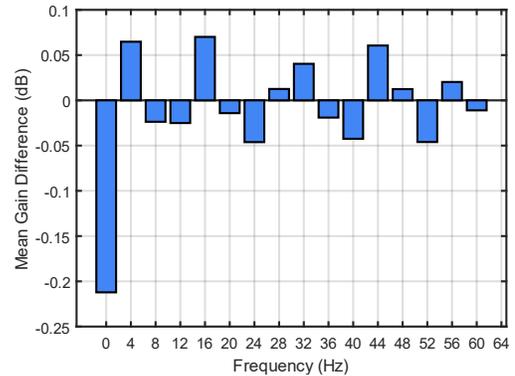

Fig. 6(b). Enhancement and Suppression of EEG Frequency Components in SincNet Filters After Training Compared to Initialization on the KUL Dataset (Two Male Speakers). Positive values indicate enhancement, negative values indicate suppression.

### C. Brain Region Correlations with Audio Envelope

In this section, we investigate the brain regions that exhibit the strongest correlation with the audio envelope within the 12 Hz to 16 Hz frequency range. We calculated the Pearson correlation coefficient between the output of the 12 Hz to 16 Hz filter in the SincNet module for each EEG channel and the corresponding audio envelope on a per-second basis. The statistical results from 18 subjects in the DTU dataset were then visualized using normalized topographical EEG maps, as shown in Fig. 7.

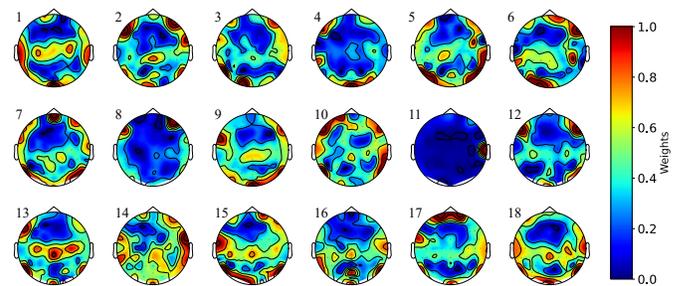

Fig. 7. Normalized topographical EEG maps depicting the Pearson correlation coefficients between EEG activity in the 12 Hz to 16 Hz frequency band and the attended audio envelope across 18 subjects in the DTU dataset. Each subplot corresponds to an individual subject, identified by their subject number in the top-left corner. Warmer colors (closer to red) indicate higher positive correlation values, while cooler colors (closer to blue) represent lower correlations.

The results identify three primary regions with significant correlations: the temporal lobe, prefrontal lobe, and parietal-occipital regions. Specifically, correlations near the left and right temporal lobes (channels T7 and T8), which encompass the primary and secondary auditory cortices, are critical for auditory attention and processing [38], [39]. The prefrontal lobe regions are associated with the maintenance of attention, selective attention, and inhibitory control, suggesting that individuals are actively concentrating on external stimuli requiring a high degree of cognitive control [40]. This suggests that the prefrontal lobe plays a crucial role in sustaining and directing attention towards relevant auditory information.

Additionally, notable correlations were observed near the parietal-occipital regions. These regions are closely related to the processing of visual information and spatial localization, reflecting their involvement in spatial awareness and perception within the environment [41].

### D. Less is More?

Building on the conclusions from the section C, we investigated whether auditory attention detection could rely on six channels located in the temporal lobe (T7, T8, FT7, FT8, TP7, TP8). To evaluate this, we compared the performance of SincAlignNet using either 6 electrodes (Model-6) or the full 64 electrodes (Model-64) on the DTU and KUL datasets. The experimental results are summarized in Table IV and V.

On the KUL dataset, Model-6 achieved the highest accuracy of 79.66 ± 20.86% with a 1-second decision window, significantly outperforming Model-64, which achieved 74.50 ± 15.99%. Similarly, on the DTU dataset, Model-6 outperformed Model-64 across multiple decision windows, particularly achieving a 1-second window accuracy of 95.03 ± 1.86% compared to Model-64's 92.98 ± 3.65%, an improvement of 2.05%. These results highlight the importance of the temporal lobe, indicating that reducing the number of electrodes has minimal impact on model performance and, in some cases, even enhance performance. This also reflects the model's stability and effectiveness.

In conclusion, the 6-electrode model based on the temporal lobe performs comparably to, and in some cases even better than, the 64-electrode model, particularly with shorter decision windows. These findings suggest that it may be possible to simplify brain-computer interface designs without significantly compromising effectiveness.

Table IV: Comparison Between 6-Electrode and 64-Electrode SincAlignNet Models for KUL Datasets.

| Decision window | Model-64 | Model-6 |
| --- | --- | --- |
| 0.5s | 68.35±21.22 | 66.47±24.96 |
| 1.0s | 74.50±15.99 | **79.66±20.86** |
| 1.5s | 78.01±18.41 | 75.79±26.69 |
| 2.0s | 72.05±18.37 | 77.89±18.27 |
| 2.5s | 74.70±17.25 | 78.70±19.51 |

Table V: Comparison Between 6-Electrode and 64-Electrode SincAlignNet Models for DTU Datasets.

| Decision window | Model-64 | Model-6 |
| --- | --- | --- |
| 0.5s | 81.66±13.94 | 84.90±15.19 |
| 1.0s | 92.98±3.65 | **95.03±1.86** |
| 1.5s | 91.56±6.85 | 88.17±8.16 |
| 2.0s | 91.78±7.39 | 92.76±8.27 |
| 2.5s | 91.70±6.23 | 88.70±9.51 |

### E. Computational Cost

We compare the training parameter counts of SincAlignNet, CNN, BIAnet, Graph EEG net and DARNet on the DTU dataset with a 1-second decision window, as summarized in Table VI. SincAlignNet employs 0.34 million parameters, significantly fewer than CNN (4.21 million) and BIAnet (3.11 million), while achieving a much higher accuracy of 92.2%, compared to 56.6% for CNN and 59.1% for BIAnet. In comparison to lightweight models such as Graph EEG Net (0.15 million) and DARNet (0.08 million), SincAlignNet uses 2.26 times and 4.25 times more parameters, respectively. However, it achieves a 43.4% improvement in accuracy over Graph EEG Net and a 33.4% improvement over DARNet. These results underscore the ability of SincAlignNet to balance parameter efficiency and high accuracy, making it a compelling choice for real-world applications where computational resources may be limited. Based on the conclusions from Sections D and E, our model could potentially be deployed on wireless ear EEG devices[42], [43], as illustrated in Fig. 8, in the future.

Table VI: Training Parameter Counts of Different Models Evaluated on DTU Dataset Using a 1-Second Decision Window.

| Model | params | Accuracy(%) |
| --- | --- | --- |
| CNN [22] | 4.21M | 56.6 |
| BIAnet [33] | 3.11M | 59.1 |
| Graph-EEG Net [34] | 0.15M | 64.3 |
| DARNet [24] | 0.08M | 69.1 |
| **SincAlignNet (ours)** | 0.34 M | 92.2 |

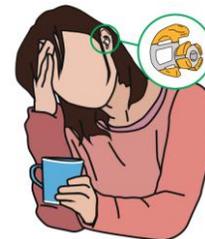

Fig. 8. Wireless ear EEG device[42].

### V. CONCLUSION

Neuroscience has provided valuable insights for designing brain-like artificial intelligence models, but methods for transferring human auditory attention abilities to neural networks have yet to be developed. In this paper, we propose SincAlignNet, a network based on SincNet and contrastive learning, designed to align audio and EEG features for auditory attention detection. The improved SincNet model simulates the brain's processing of audio, while contrastive learning helps us understand the brain's behavior patterns during auditory attention. We evaluated the model using rigorous cross-trial testing on two datasets. The results show that SincAlignNet achieves state-of-the-art decoding accuracies of 78.3% and 92.2% on the KUL and DTU datasets, respectively, with a 1-second decision window. We also found that using only six electrodes in the temporal lobe provides performance comparable to, or even better than, using 64 electrodes, offering potential for future neuro-guided hearing aids.

However, there are areas for improvement. Auditory attention selection typically involves interactions across multiple brain regions, which could be better modeled using graph neural networks in the future. Additionally, considering frequency features alone may not be sufficient; incorporating more contextual information and the content of the audio could further enhance model performance. The model's generalizability still requires further exploration, particularly in scenarios with multiple speakers and background noise. In future research, we plan to further explore and refine the SincAlignNet model to improve its generalizability and robustness.

## ACKNOWLEDGMENT

We would like to express our gratitude to Jiangsu ADE Intelligent Technology Co., Ltd. for providing the A100 server computing platform that was essential for conducting this research. Additionally, we extend our sincere thanks to the SCCN Lab at the University of California, San Diego, and especially to Professor Tzyy-Ping Jung for his valuable suggestions and insightful feedback that significantly improved the quality of this paper.